\documentclass[letterpaper,10pt]{article} 
\usepackage{osameet3} %% use version 3 for proper copyright statement

\usepackage{amsmath,amssymb}
\usepackage[colorlinks=true,bookmarks=false,citecolor=blue,urlcolor=blue]{hyperref} %pdflatex

\begin{document}

\title{
A Molecular-MNIST Dataset for Machine Learning Study on Diffraction Imaging and Microscopy}

\author{Yan Zhang$^1$, Steve Farrell$^1$, Michael Crowley$^2$, Lee Makowski$^3$, Jack Deslippe$^1$}
\address{$^1$NERSC, Lawrence Berkeley National Laboratory, Berkeley, CA\\$^2$Bioscience Division, National Renewable Energy Laboratory, Golden, CO\\$^3$Department of Bioengineering, Northeastern University, Boston, MA}
\email{yanzhang@lbl.gov}

%% Uncomment the following line to override copyright year from the default current year.
\copyrightyear{2020}

\begin{abstract}
An image dataset of 10 different size molecules, where each molecule has 2,000 structural variants, is generated from the 2D cross-sectional projection of Molecular Dynamics trajectories. The purpose of this dataset is to provide a benchmark dataset for the increasing need of machine learning, deep learning and image processing on the study of scattering, imaging and microscopy.

\end{abstract}


\section{Introduction}
Many imaging techniques require phase retrieval, such as coherent diffraction imaging, electron microscopy and etc. These iterative optimization algorithms are computational expensive and difficult to converge. Unlike iterative optimization methods, supervised machine learning using two stage training-testing becomes a great advantage for fast real-time inference since the most expensive computations are performed during training. Deep Learning plays a very important role tackling these type of problems but requires large dataset to train the multi-layer model parameters of the network \cite{alex}. Here, we are interested in creating a molecular image dataset including shape images from real space and diffraction patterns from reciprocal space for machine learning practices. We call this dataset Molecular-MNIST because it consists 10 different size of molecules where each molecule has 2,000 structural variants - in an analogy of the famous 10-digit hand-written dataset MNIST \cite{mnist}.

\section{Molecular-MNIST Dataset}

\subsection{3D Molecular Structures}
The molecular structures were simulated using Charmm generating 200 $\AA$ long $I\beta$ crystalline structure of various shape and size. It consists molecules of diamond shape of 2 $\times$ 2, 3 $\times$ 3, 4 $\times$ 4, 5 $\times$ 5, 6 $\times$ 6, 7 $\times$ 7, 8 $\times$ 8, 9 $\times$ 9 -chains, and hexagonal shape of 24 and 36 -chains. All of the molecules were simulated under structural dynamics in water buffer until equilibrium.

\subsection{2D Cross-sectional Shape Images and Diffraction Patterns}
\begin{figure*}[htb]
\centering

\begin{minipage}[b]{0.09\linewidth}
  \centering
  \centerline{\includegraphics[width=2.cm]{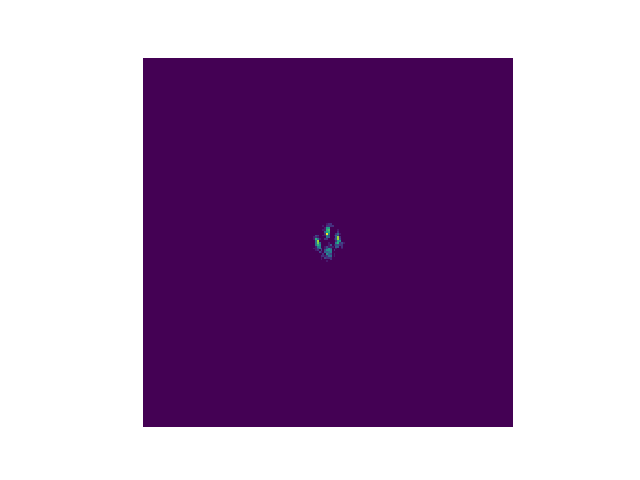}}
%  \vspace{1.5cm}
  %\centerline{(b) Results 3}\medskip
\end{minipage}
%\hfill
\begin{minipage}[b]{0.09\linewidth}
  \centering
  \centerline{\includegraphics[width=2.cm]{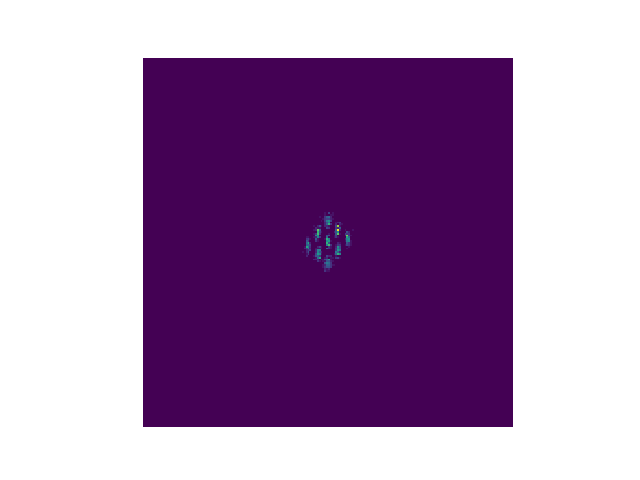}}
%  \vspace{1.5cm}
  %\centerline{(c) Result 4}\medskip
\end{minipage}
%\hfill
\begin{minipage}[b]{0.09\linewidth}
  \centering
  \centerline{\includegraphics[width=2.cm]{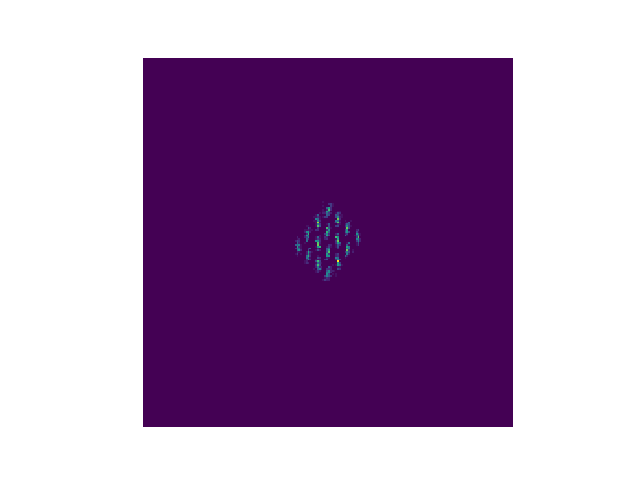}}
%  \vspace{1.5cm}
  %\centerline{(c) Result 4}\medskip
\end{minipage}
%\hfill
\begin{minipage}[b]{0.09\linewidth}
  \centering
  \centerline{\includegraphics[width=2.cm]{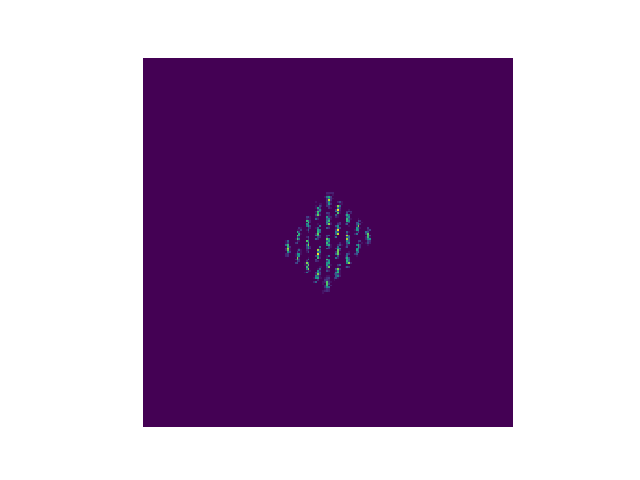}}
%  \vspace{1.5cm}
  %\centerline{(c) Result 4}\medskip
\end{minipage}
%\hfill
\begin{minipage}[b]{0.09\linewidth}
  \centering
  \centerline{\includegraphics[width=2.cm]{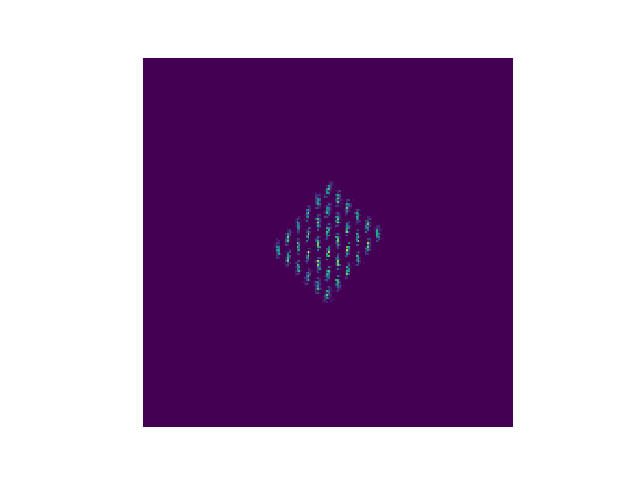}}
%  \vspace{1.5cm}
  %\centerline{(c) Result 4}\medskip
\end{minipage}
%\hfill
\begin{minipage}[b]{0.09\linewidth}
  \centering
  \centerline{\includegraphics[width=2.cm]{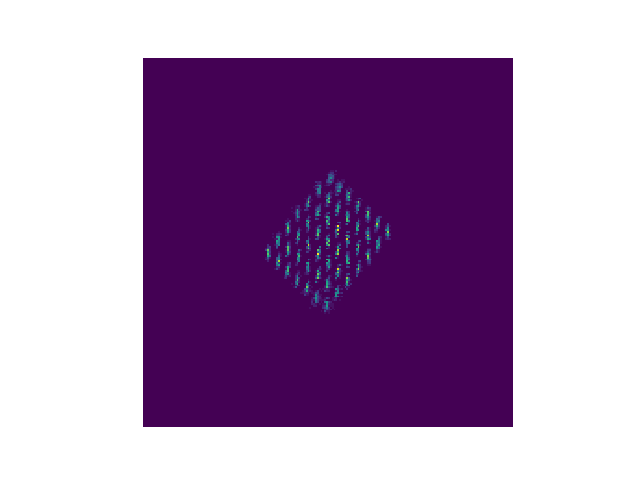}}
%  \vspace{1.5cm}
  %\centerline{(c) Result 4}\medskip
\end{minipage}
%\hfill
\begin{minipage}[b]{0.09\linewidth}
  \centering
  \centerline{\includegraphics[width=2.cm]{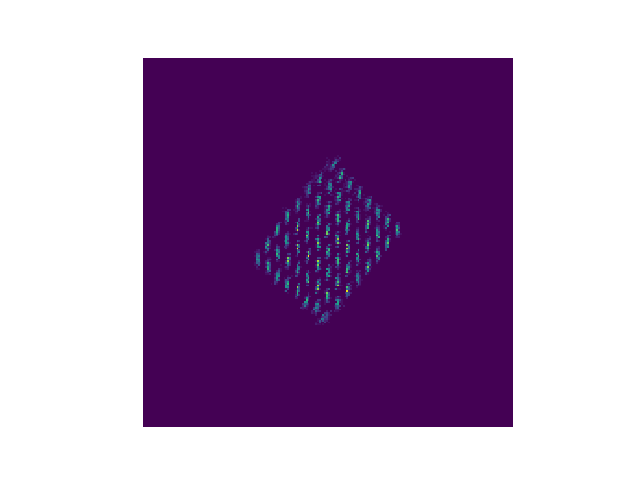}}
%  \vspace{1.5cm}
  %\centerline{(c) Result 4}\medskip
\end{minipage}
%\hfill
\begin{minipage}[b]{0.09\linewidth}
  \centering
  \centerline{\includegraphics[width=2.cm]{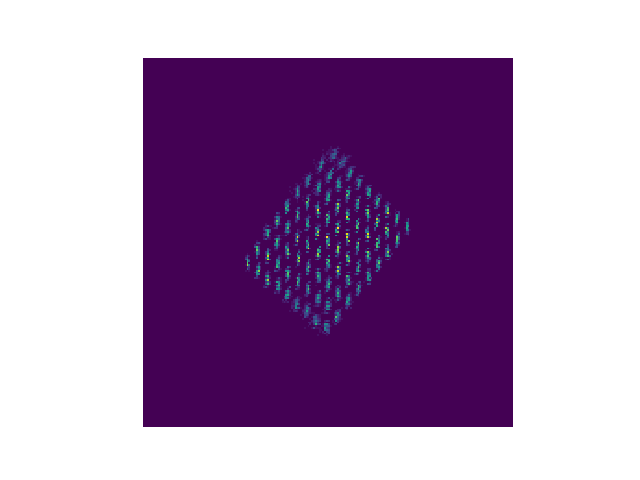}}
%  \vspace{1.5cm}
  %\centerline{(c) Result 4}\medskip
\end{minipage}
%\hfill
\begin{minipage}[b]{0.09\linewidth}
  \centering
  \centerline{\includegraphics[width=2.cm]{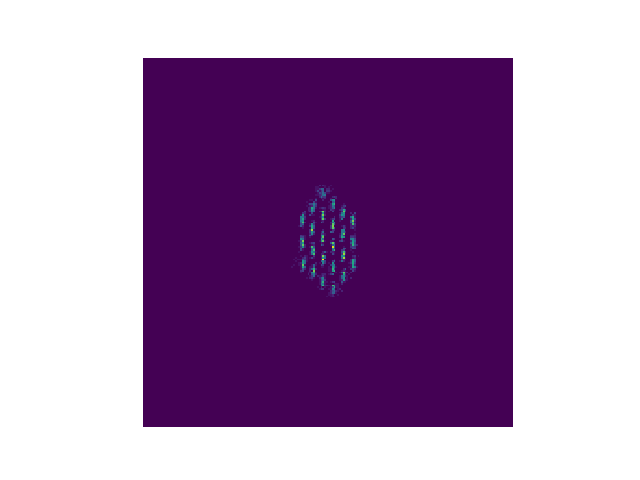}}
%  \vspace{1.5cm}
  %\centerline{(c) Result 4}\medskip
\end{minipage}
%\hfill
\begin{minipage}[b]{0.09\linewidth}
  \centering
  \centerline{\includegraphics[width=2.cm]{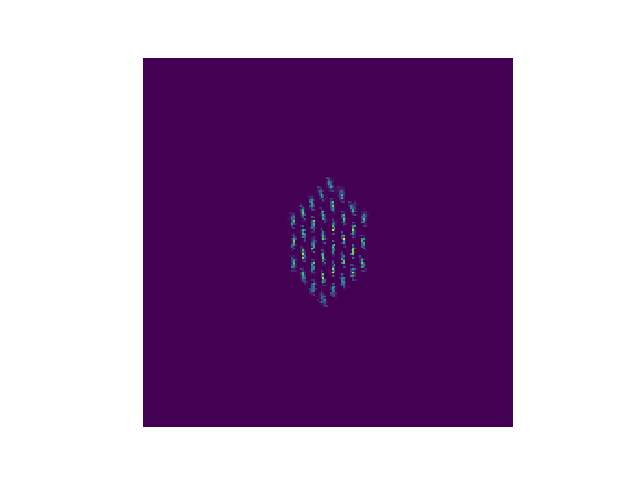}}
%  \vspace{1.5cm}
  %\centerline{(c) Result 4}\medskip
\end{minipage}

\begin{minipage}[b]{0.09\linewidth}
  \centering
  \centerline{\includegraphics[width=2.cm]{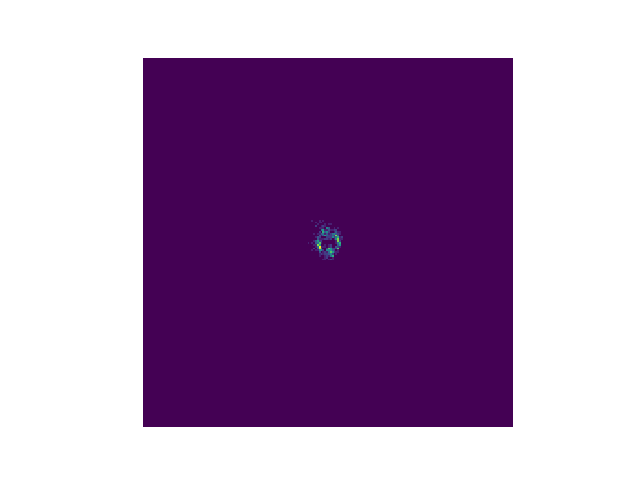}}
%  \vspace{1.5cm}
  %\centerline{(b) Results 3}\medskip
\end{minipage}
%\hfill
\begin{minipage}[b]{0.09\linewidth}
  \centering
  \centerline{\includegraphics[width=2.cm]{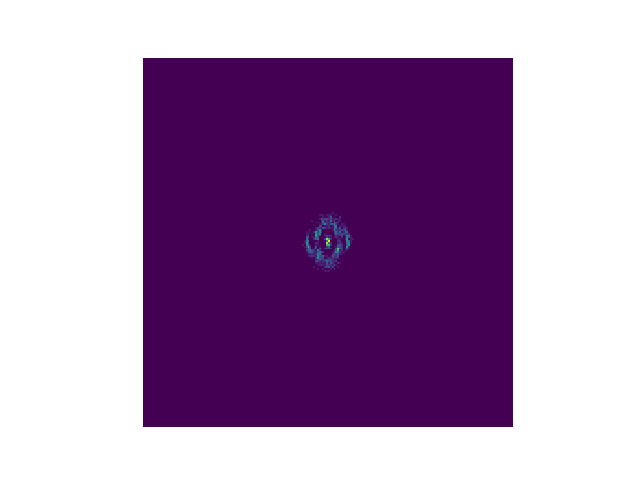}}
%  \vspace{1.5cm}
  %\centerline{(c) Result 4}\medskip
\end{minipage}
%\hfill
\begin{minipage}[b]{0.09\linewidth}
  \centering
  \centerline{\includegraphics[width=2.cm]{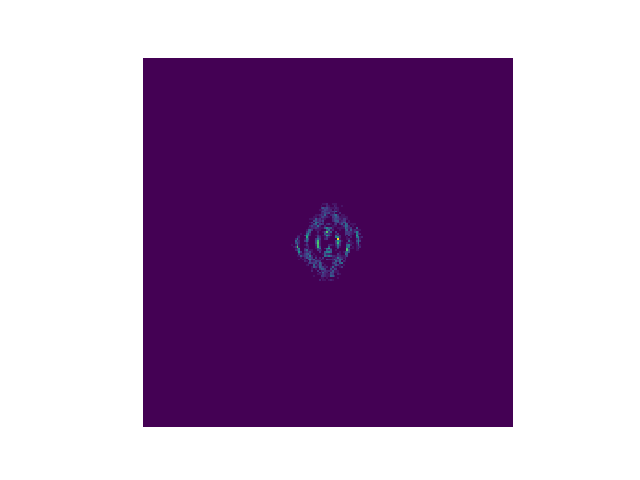}}
%  \vspace{1.5cm}
  %\centerline{(c) Result 4}\medskip
\end{minipage}
%\hfill
\begin{minipage}[b]{0.09\linewidth}
  \centering
  \centerline{\includegraphics[width=2.cm]{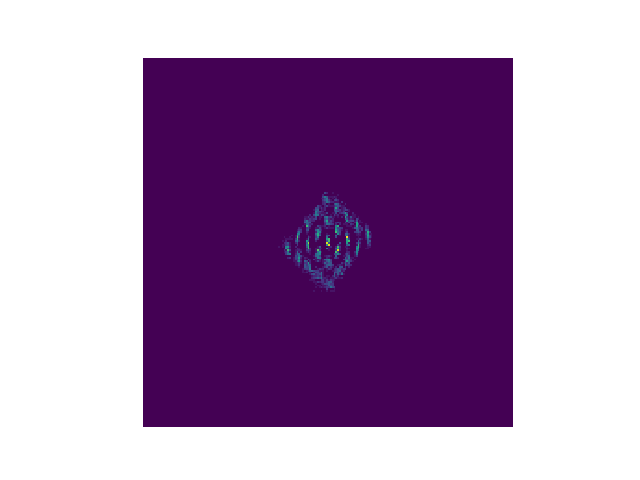}}
%  \vspace{1.5cm}
  %\centerline{(c) Result 4}\medskip
\end{minipage}
%\hfill
\begin{minipage}[b]{0.09\linewidth}
  \centering
  \centerline{\includegraphics[width=2.cm]{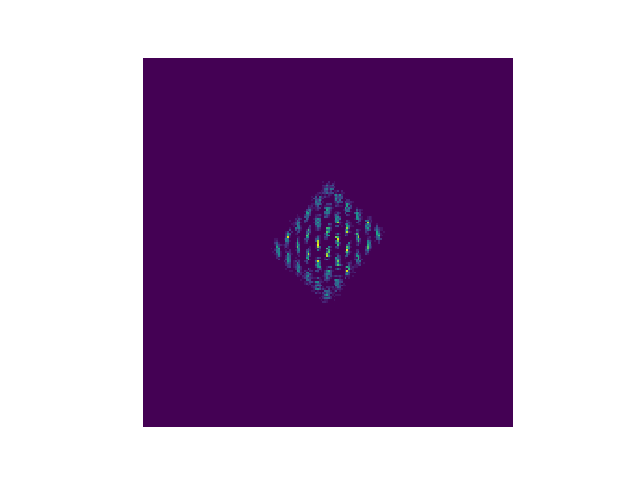}}
%  \vspace{1.5cm}
  %\centerline{(c) Result 4}\medskip
\end{minipage}
%\hfill
\begin{minipage}[b]{0.09\linewidth}
  \centering
  \centerline{\includegraphics[width=2.cm]{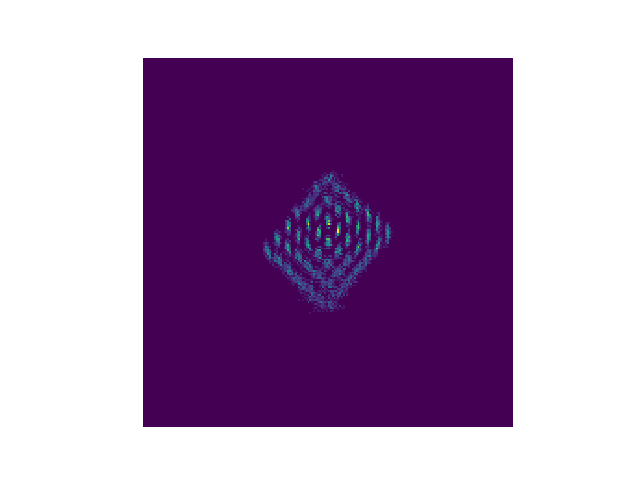}}
%  \vspace{1.5cm}
  %\centerline{(c) Result 4}\medskip
\end{minipage}
%\hfill
\begin{minipage}[b]{0.09\linewidth}
  \centering
  \centerline{\includegraphics[width=2.cm]{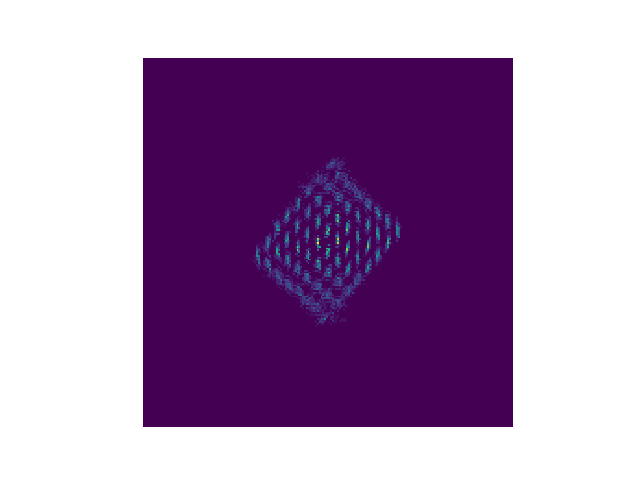}}
%  \vspace{1.5cm}
  %\centerline{(c) Result 4}\medskip
\end{minipage}
%\hfill
\begin{minipage}[b]{0.09\linewidth}
  \centering
  \centerline{\includegraphics[width=2.cm]{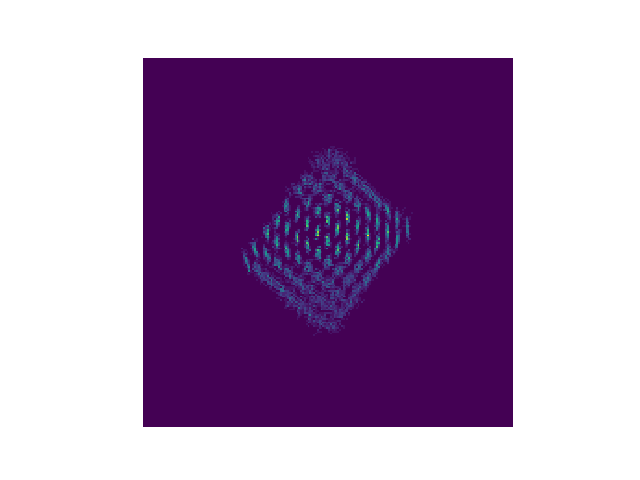}}
%  \vspace{1.5cm}
  %\centerline{(c) Result 4}\medskip
\end{minipage}
%\hfill
\begin{minipage}[b]{0.09\linewidth}
  \centering
  \centerline{\includegraphics[width=2.cm]{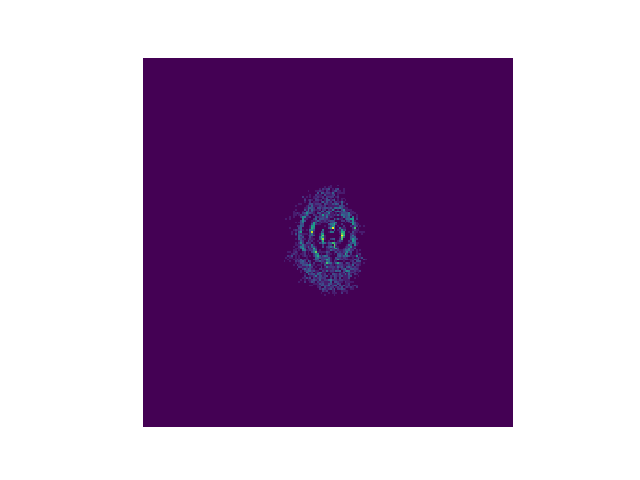}}
%  \vspace{1.5cm}
  %\centerline{(c) Result 4}\medskip
\end{minipage}
%\hfill
\begin{minipage}[b]{0.09\linewidth}
  \centering
  \centerline{\includegraphics[width=2.cm]{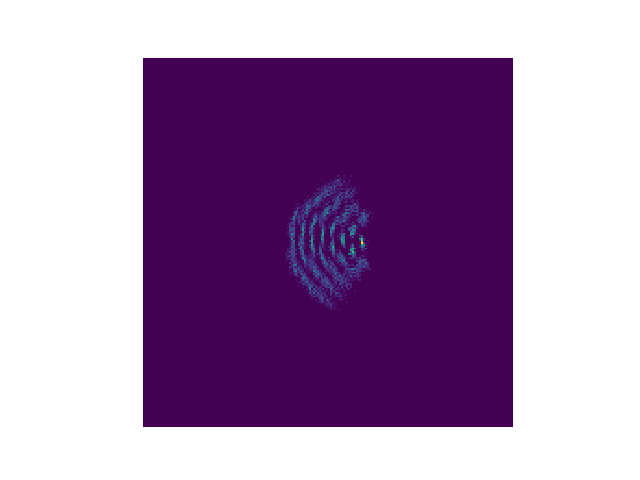}}
%  \vspace{1.5cm}
  %\centerline{(c) Result 4}\medskip
\end{minipage}

\caption{Shape images in Molecular-MNIST dataset. From left to right: different shapes in the order of 4, 9, 16, 25, 36, 49, 64, 81 diamond shape and 24, 36 hexagonal shape. Top row has the least twisted variation, bottom row has the most twisted variation.}
\label{fig:shape}
\end{figure*}

\begin{figure*}[htb]
\centering

\begin{minipage}[b]{0.09\linewidth}
  \centering
  \centerline{\includegraphics[width=2.cm]{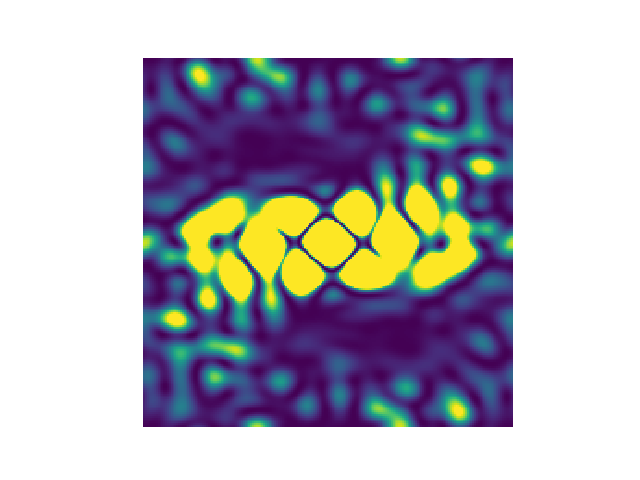}}
%  \vspace{1.5cm}
  %\centerline{(b) Results 3}\medskip
\end{minipage}
%\hfill
\begin{minipage}[b]{0.09\linewidth}
  \centering
  \centerline{\includegraphics[width=2.cm]{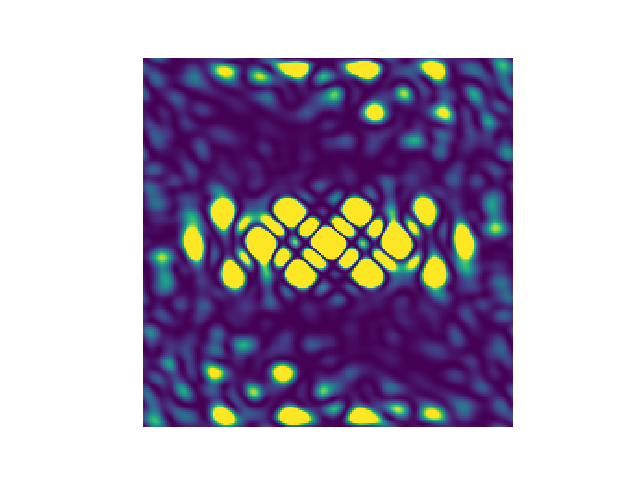}}
%  \vspace{1.5cm}
  %\centerline{(c) Result 4}\medskip
\end{minipage}
%\hfill
\begin{minipage}[b]{0.09\linewidth}
  \centering
  \centerline{\includegraphics[width=2.cm]{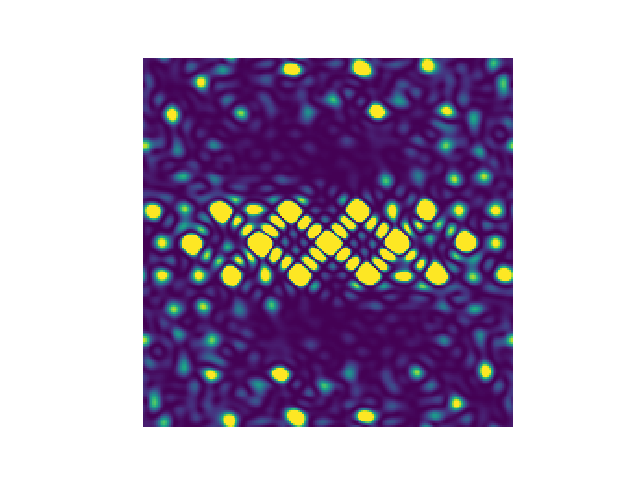}}
%  \vspace{1.5cm}
  %\centerline{(c) Result 4}\medskip
\end{minipage}
%\hfill
\begin{minipage}[b]{0.09\linewidth}
  \centering
  \centerline{\includegraphics[width=2.cm]{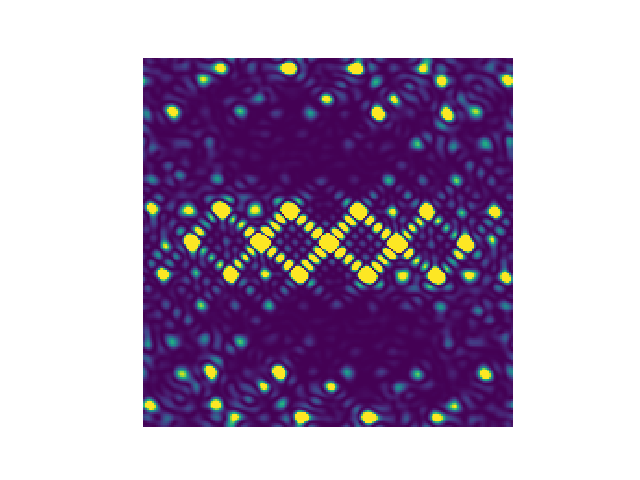}}
%  \vspace{1.5cm}
  %\centerline{(c) Result 4}\medskip
\end{minipage}
%\hfill
\begin{minipage}[b]{0.09\linewidth}
  \centering
  \centerline{\includegraphics[width=2.cm]{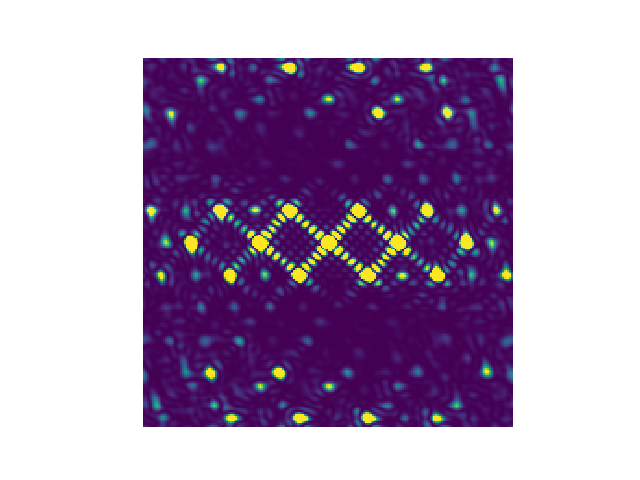}}
%  \vspace{1.5cm}
  %\centerline{(c) Result 4}\medskip
\end{minipage}
%\hfill
\begin{minipage}[b]{0.09\linewidth}
  \centering
  \centerline{\includegraphics[width=2.cm]{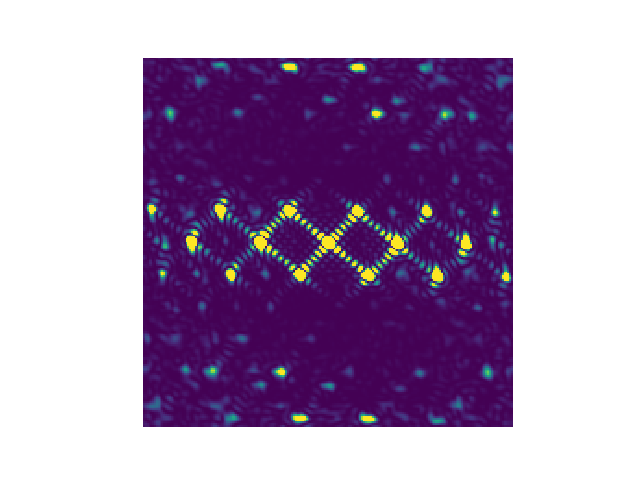}}
%  \vspace{1.5cm}
  %\centerline{(c) Result 4}\medskip
\end{minipage}
%\hfill
\begin{minipage}[b]{0.09\linewidth}
  \centering
  \centerline{\includegraphics[width=2.cm]{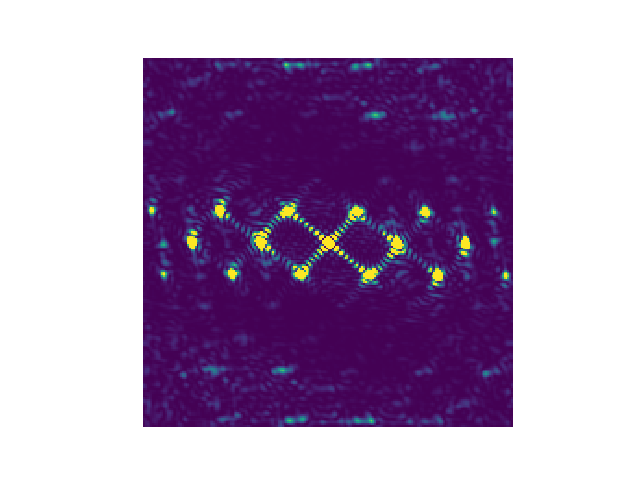}}
%  \vspace{1.5cm}
  %\centerline{(c) Result 4}\medskip
\end{minipage}
%\hfill
\begin{minipage}[b]{0.09\linewidth}
  \centering
  \centerline{\includegraphics[width=2.cm]{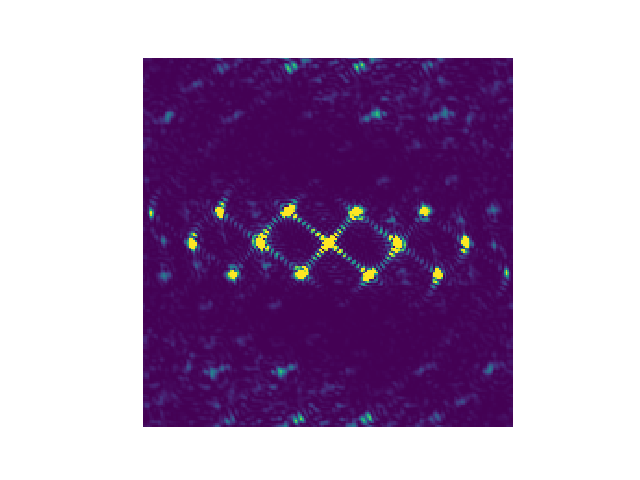}}
%  \vspace{1.5cm}
  %\centerline{(c) Result 4}\medskip
\end{minipage}
%\hfill
\begin{minipage}[b]{0.09\linewidth}
  \centering
  \centerline{\includegraphics[width=2.cm]{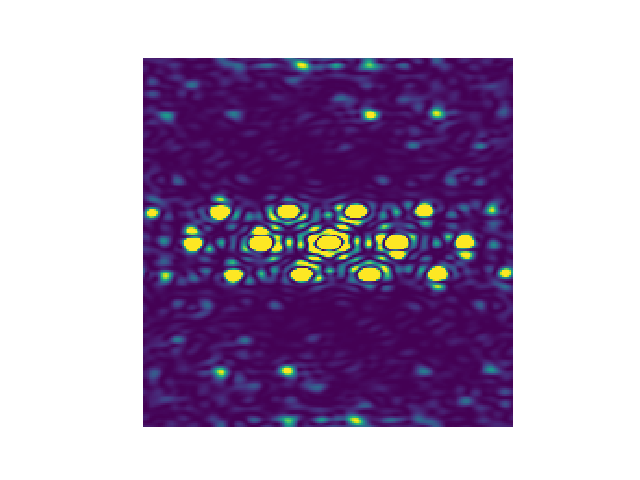}}
%  \vspace{1.5cm}
  %\centerline{(c) Result 4}\medskip
\end{minipage}
%\hfill
\begin{minipage}[b]{0.09\linewidth}
  \centering
  \centerline{\includegraphics[width=2.cm]{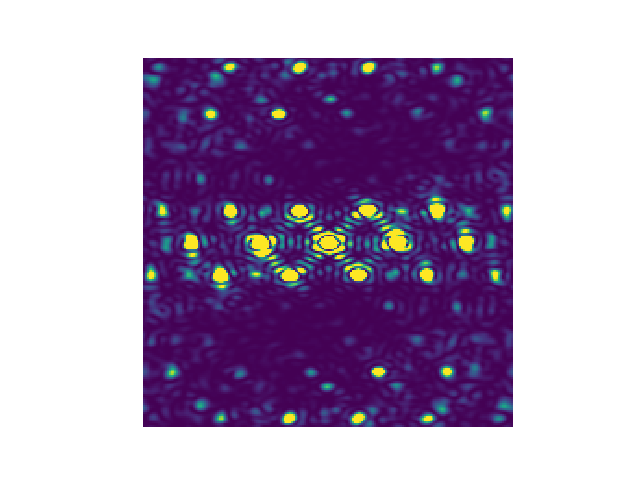}}
%  \vspace{1.5cm}
  %\centerline{(c) Result 4}\medskip
\end{minipage}

\begin{minipage}[b]{0.09\linewidth}
  \centering
  \centerline{\includegraphics[width=2.cm]{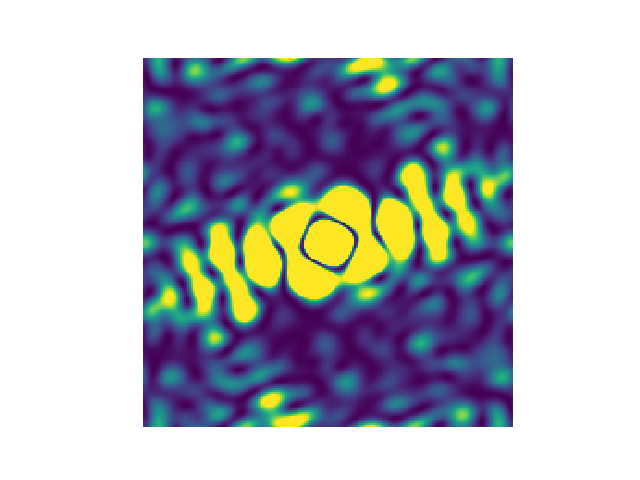}}
%  \vspace{1.5cm}
  %\centerline{(b) Results 3}\medskip
\end{minipage}
%\hfill
\begin{minipage}[b]{0.09\linewidth}
  \centering
  \centerline{\includegraphics[width=2.cm]{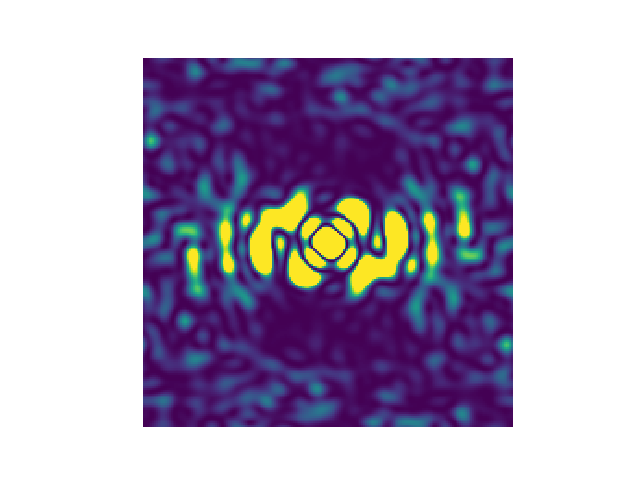}}
%  \vspace{1.5cm}
  %\centerline{(c) Result 4}\medskip
\end{minipage}
%\hfill
\begin{minipage}[b]{0.09\linewidth}
  \centering
  \centerline{\includegraphics[width=2.cm]{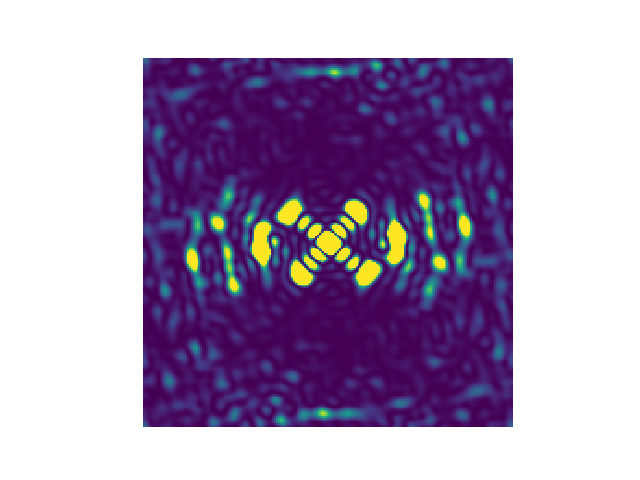}}
%  \vspace{1.5cm}
  %\centerline{(c) Result 4}\medskip
\end{minipage}
%\hfill
\begin{minipage}[b]{0.09\linewidth}
  \centering
  \centerline{\includegraphics[width=2.cm]{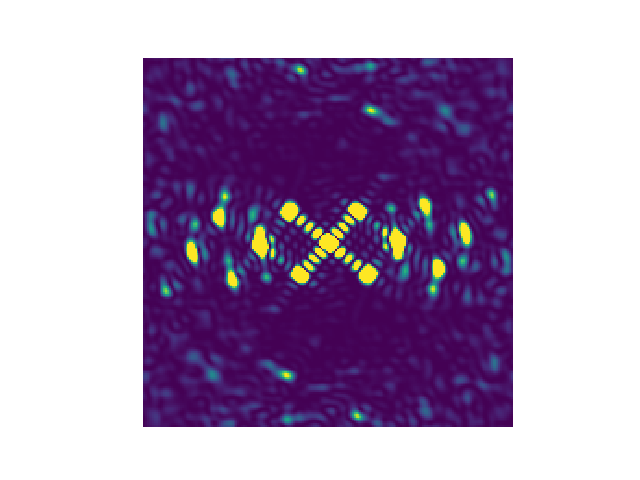}}
%  \vspace{1.5cm}
  %\centerline{(c) Result 4}\medskip
\end{minipage}
%\hfill
\begin{minipage}[b]{0.09\linewidth}
  \centering
  \centerline{\includegraphics[width=2.cm]{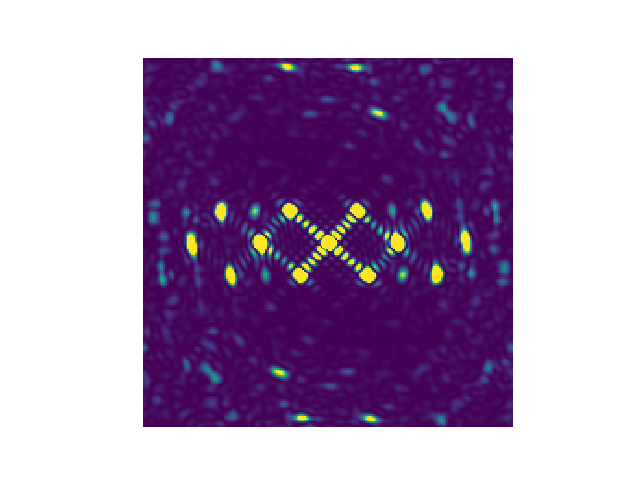}}
%  \vspace{1.5cm}
  %\centerline{(c) Result 4}\medskip
\end{minipage}
%\hfill
\begin{minipage}[b]{0.09\linewidth}
  \centering
  \centerline{\includegraphics[width=2.cm]{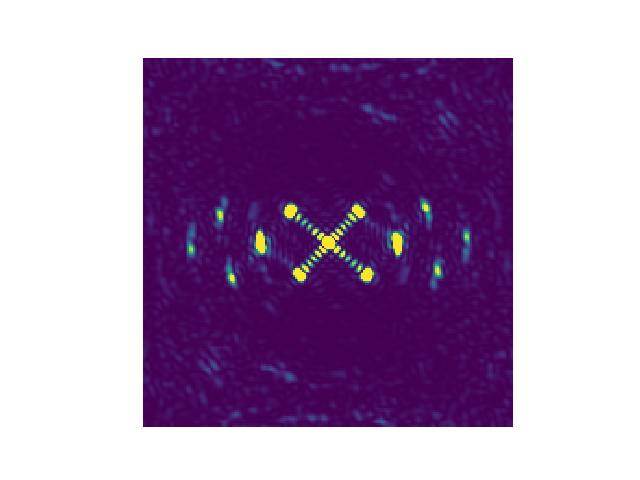}}
%  \vspace{1.5cm}
  %\centerline{(c) Result 4}\medskip
\end{minipage}
%\hfill
\begin{minipage}[b]{0.09\linewidth}
  \centering
  \centerline{\includegraphics[width=2.cm]{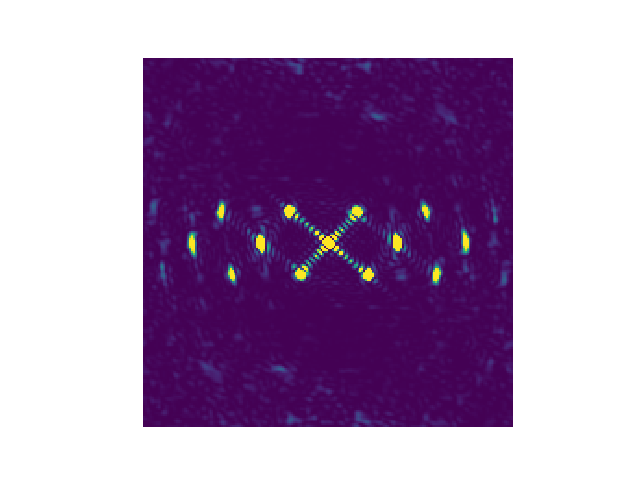}}
%  \vspace{1.5cm}
  %\centerline{(c) Result 4}\medskip
\end{minipage}
%\hfill
\begin{minipage}[b]{0.09\linewidth}
  \centering
  \centerline{\includegraphics[width=2.cm]{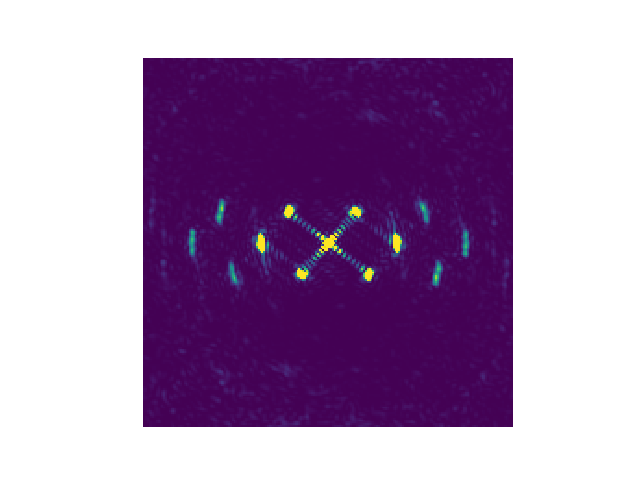}}
%  \vspace{1.5cm}
  %\centerline{(c) Result 4}\medskip
\end{minipage}
%\hfill
\begin{minipage}[b]{0.09\linewidth}
  \centering
  \centerline{\includegraphics[width=2.cm]{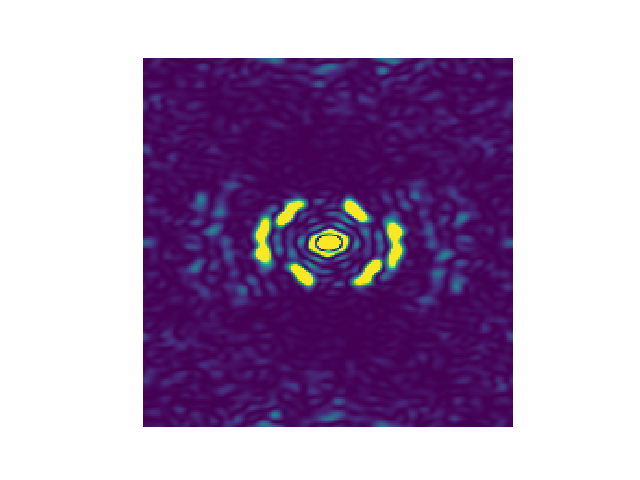}}
%  \vspace{1.5cm}
  %\centerline{(c) Result 4}\medskip
\end{minipage}
%\hfill
\begin{minipage}[b]{0.09\linewidth}
  \centering
  \centerline{\includegraphics[width=2.cm]{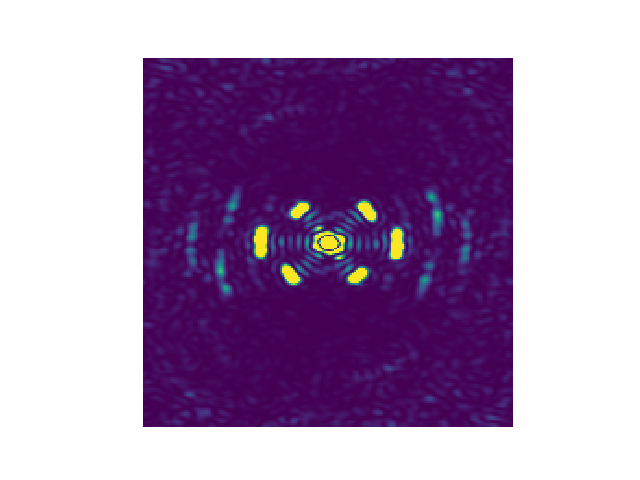}}
%  \vspace{1.5cm}
  %\centerline{(c) Result 4}\medskip
\end{minipage}

\caption{Diffraction patterns in Molecular-MNIST dataset. From left to right: patterns from different shapes in the order of 4, 9, 16, 25, 36, 49, 64, 81 diamond and 24, 36 hexagonal. Top row has the patterns from least twisted variation, bottom row has the patterns from most twisted variation.}
\label{fig:cdi}
\end{figure*}

We projected the 3D atomic distribution of these molecular structures onto a 2D cross-sectional plane to create 2D shape images as shown in Fig.\ref{fig:shape}. The $(x,y)$ positions of all the atoms in the 3D coordinates were projected onto a 2D grid using $histogram2d$ functions multiplied by their electron densities \cite{yan}. 
%The axis of the 2D image are $\lfloor -max(x_{max}, y_{max})-\Delta \rfloor$ and $\lfloor max(x_{max}, y_{max})+\Delta \rfloor$ for both horizontal and vertical and 
The image resolution was scaled to $200 \times 200$ which is also changeable for higher resolution if necessary.
We took the 2D Fourier Transform of the cross-sectional shape images and the diffraction patterns were the square of amplitude of the FFT results as shown in Fig.\ref{fig:cdi}. 
%\subsection{Diffraction Patterns}

%Researchers could use this dataset to train a deep neural network that performs the diffraction inversion without using the phase information. Noted that some researchers are also interested in estimating the phase through diffraction, which should be trainable as well.

\subsection{Data Distribution}
In order to visualize the data distribution, we computed and scattered the log-sum of low frequency intensities versus the log-sum of high frequency intensities in a 2D-plot as shown in Fig \ref{fig:vis} left.

\section{Use cases and Conclusion}
\begin{figure*}[htb]

\begin{minipage}[b]{0.5\linewidth}
  %\centering
  \centerline{\includegraphics[width=4.5cm]{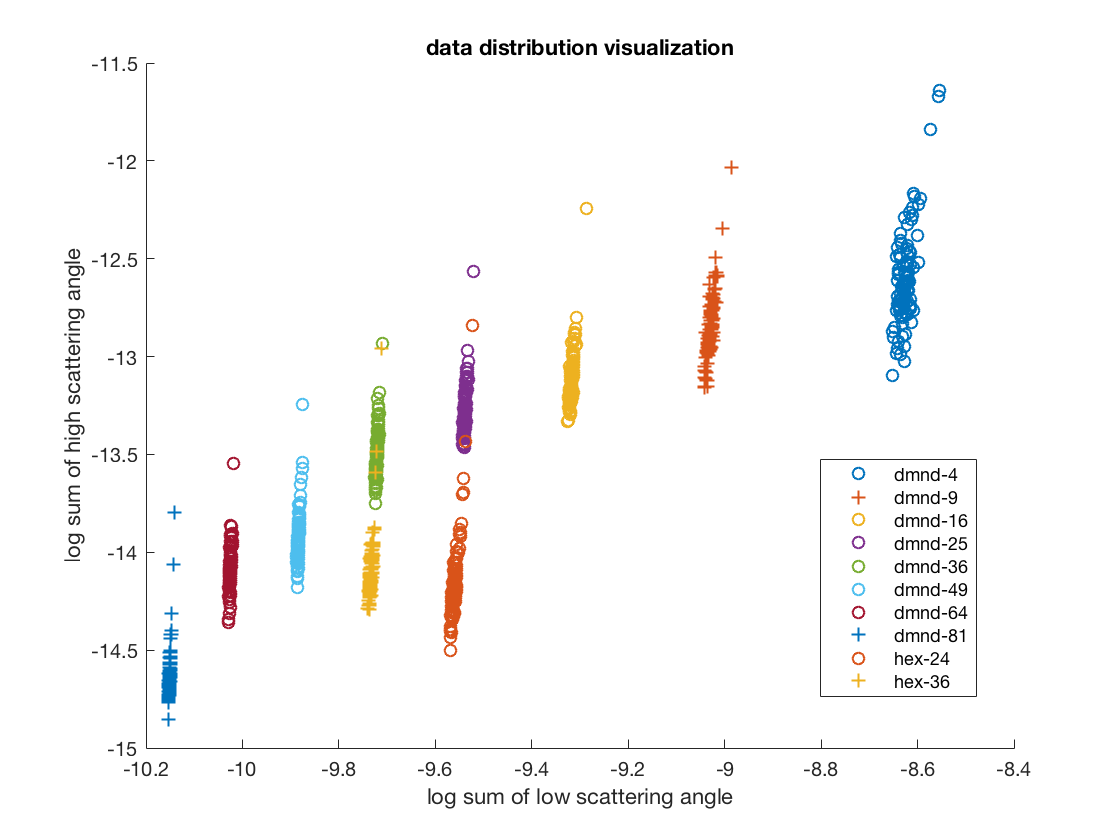}}
  \vspace{-3.0cm}
  \hspace{7.0cm}
  \centerline{\includegraphics[width=6.0cm]{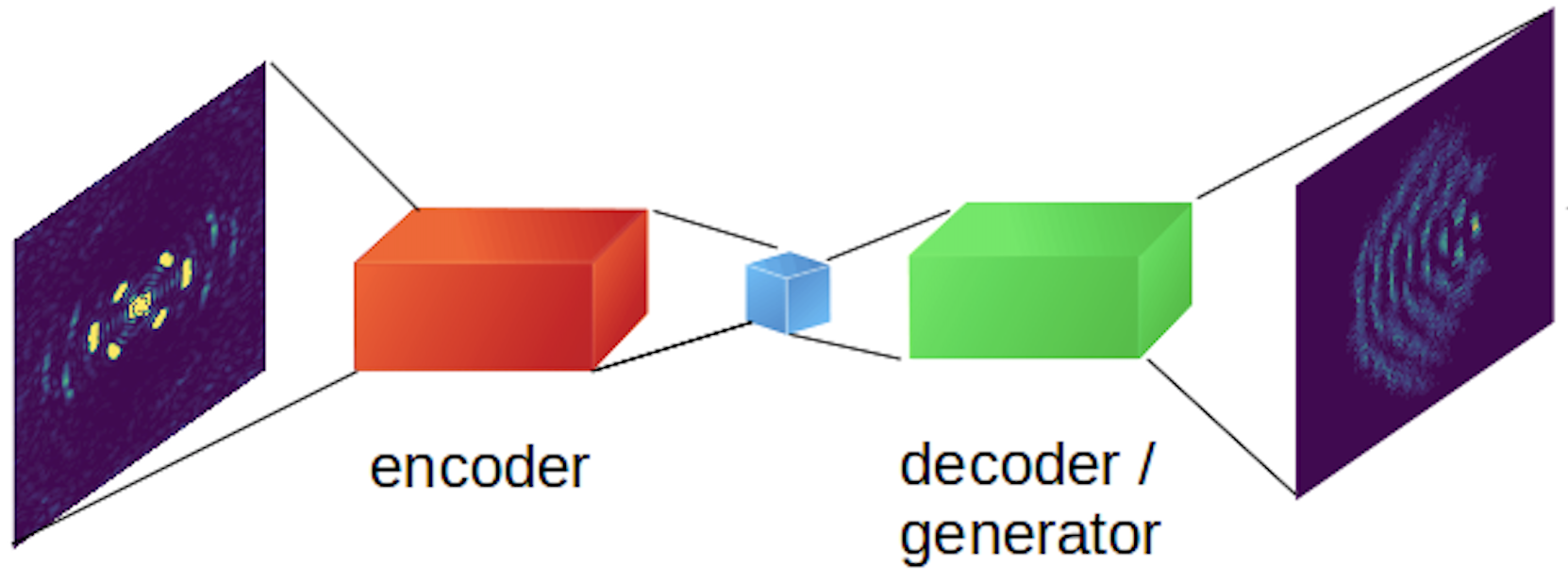}}

%  \vspace{2.0cm}
  %\centerline{(a) Result 1}\medskip
\end{minipage}

%\begin{minipage}[b]{0.8\linewidth}
%  \centering
%  \centerline{\includegraphics[width=3.0cm]{vis/autoencoder.png}}
%  \vspace{2.0cm}
  %\centerline{(a) Result 1}\medskip
%\end{minipage}

\vspace{0.8cm}

\caption{Left: Data Distribution using log-sum of low frequency vs. log-sum of high frequency intensities. Right: Convolutional encoder-decoder network for diffraction inversion, a use case of the Molecular-MNIST dataset.}
\label{fig:vis}
\end{figure*}

Here, we used a convolutional encoder-decoder network for diffraction inversion as shown in Fig.\ref{fig:vis} right. The encoder and decoder have symmetric architectures with each other and use only 3 $\times$ 3 kernels, similar with VGG-Net \cite{vgg}. The encoder has 4 convolutional layers each with 8, 16, 32 and 64 kernels, followed by batch-normalization and ReLU activation. The decoder, or a deconvolutional network, takes the 64 features from the output of the encoder and passes it to its reversed architecture of the encoder. After fully trained, this network was capable to inference molecular shape without phase retrieval. We believe this Molecular-MNIST dataset has a huge potential by serving as a benchmark dataset for many deep learning studies. Download Molecular-MNIST at: https://www.dropbox.com/sh/t8unmhfm6epemlq/AAA6TSFNNSRe-e6JWlOZH5xra?dl=0

%%%%%%%%%%%%%%%%%%%%%%%%%%%%%%%%%%%%%%%%%%%%%%%%%%%%%%%%%%%%%%%%%%%%%%%%%%%%%%%%%%%%%%%%%%%%%

\vspace{-0.5cm}
%%%%%%%%%%%%%%%%%%%%%%%%%%%%%%%%%%%%%%%%%%%%%%%%%%%%%%%%%%%%%%%%%%%%%%%%%%%%%%%%%%%%%%%%%%%%%

%%%%%%%%%%%%%%%%%%%%%%%%%%%%%%%%%%%%%%%%%%%%%%%%%%%%%%%%%%%%%%%%%%%%%%%%%%%%%%%%%%%%%%%%%%%%%

\vspace{-1em}

%%%%%%%%%%%%%%%%%%%%%%%%%%%%%%%%%%%%%%%%%%%%%%%%%%%%%%%%%%%%%%%%%%%%%%%%%%%%%%%%%%%%%%%%%%%%%

\section{Acknowledgement}
This research used resources of the National Energy Research Scientific Computing Center (NERSC), a U.S. Department of Energy Office of Science User Facility operated under Contract No. DE-AC02-05CH11231.

\end{document}